\documentclass[reprint,prl,superscriptaddress]{revtex4-1}	
	\usepackage{amsmath}
	\usepackage{makeidx}
	\usepackage{amsfonts}
	\usepackage{natbib}
	\usepackage[ansinew]{inputenc}
	\usepackage[usenames,dvipsnames]{pstricks}
	\usepackage{subfigure}
	\usepackage{epsfig}
	\usepackage{ulem}
	\usepackage{color,soul}
	\usepackage{pst-grad} 
	\usepackage{pst-plot} 
	\usepackage[colorlinks,hyperindex]{hyperref}
	\hypersetup
	{
		colorlinks,%
		citecolor=black,%
		linkcolor=black,%
		urlcolor=black,%
	}

\begin{document}
\title{Stiffness heterogeneity of small viral capsids}
\author{Lucas Menou$^*$}
\affiliation{Univ Lyon, Ens de Lyon, CNRS, Laboratoire de Physique, F-69342 Lyon, France}
\author{Yeraldinne Carrasco Salas$^*$}
\affiliation{Univ Lyon, Ens de Lyon, CNRS, Laboratoire de Physique, F-69342 Lyon, France}
\author{Lauriane Lecoq}
\affiliation{Institut de Biologie et Chimie des Prot\'eines, University of Lyon 1, Lyon, France}
\author{Anna Salvetti}
\affiliation{International Center for Research in Infectiology (CIRI), INSERM U111, CNRS UMR 5308, Lyon, France}
\author{Cendrine Faivre Moskalenko}
\affiliation{Univ Lyon, Ens de Lyon, CNRS, Laboratoire de Physique, F-69342 Lyon, France}
\author{Martin Castelnovo}
\affiliation{Univ Lyon, Ens de Lyon, CNRS, Laboratoire de Physique, F-69342 Lyon, France}

\begin{abstract}
Nanoindentation of viral capsids provides an efficient tool in order to probe their elastic properties. We investigate in the present work the various sources of stiffness heterogeneity as observed in Atomic Force Microscopy (AFM) experiments. By combining experimental results with both numerical and analytical modeling, we first show that for small viruses a position-dependent stiffness is observed. This effect is strong and has not been properly taken into account previously. Moreover, we show that a geometrical model is able to reproduce this effect quantitatively. 
Our work suggests alternative ways of measuring stiffness heterogeneities on small viral capsids. This is illustrated on two different viral capsids: Adeno Associated Virus serotype 8 (AAV8) and Hepatitis B Virus (HBV with T=4). We discuss our results in the light of continuous elasticity modeling.
\end{abstract}
\date{\today}
\maketitle

\section{Introduction}
The elastic properties of viral capsids can be inferred using nanoindentation experiments. Viral capsids' sizes ranging from tens to hundreds of nanometers, these experiments are usually performed with an Atomic Force Microscope (AFM), whose operating size range is appropriate \cite{zandi2020}. Capsids of different virus families have been probed this way in several works, exhibiting various interesting features \cite{ivanovska2004,michel2006,carrasco2006,mateu2012,mateu2013,zeng2017,zeng2017_2,martingonzalez2021}. Depending on the identity of capsids and the amplitude of indenting force, one can observe elastic behavior with a linear or non-linear response, plastic behavior, and eventually mechanical rupture. From the biological point of view, all these observations are relevant since they address the global stability of viruses, as well as their deformability, which are important physical information to fully understand the various steps of their replication cycle. 

Most past works dealing with nanoindentation of capsids have focused on their linear response. This linear regime allows defining  the \textit{stiffness} effectively as the ratio between the force applied and the indentation. Typical stiffness ranges from  0.04 $N.m^{-1}$ for the Influenza virus \cite{eghiaian2009,mateu2012} to 3 $N.m^{-1}$  for the immature particle of HIV-1 in the presence of envelope glycoproteins for example \cite{kol2007}. From a structural perspective, viruses are composed of mainly nucleic acids packaged inside self-assembled protein shells and eventually a surrounding lipid bilayer or membrane. For most viral capsids, the self-assembly involves multiple copies of identical proteins. Yet, the topology of a closed shell imposes that the environment at the scale of a single protein is not unique: at exactly twelve locations on the surface of the virus, proteins belong to a \textit{capsomer} or face that is pentameric (five proteins), while they belong to hexameric capsomers (six proteins) anywhere else. The precise location of pentamers determines the global shape of the viral capsid \cite{castelnovo2017}. It can be either regular with icosahedral symmetry, irregular and elongated. Focusing on icosahedral viruses, the three main symmetries (two-fold, three-fold and five-fold) strongly suggest that elastic properties of capsids are not uniform \cite{aznar2012}.

Heterogeneities in stiffness measurements have been observed on some viruses, such as the MVM parvovirus \cite{carrasco2006}, or bacteriophages $\phi 29$  \cite{ivanovska2004}. In the former case, heterogeneities have been attributed to the difference in virus orientation during the adhesion on the substrate, offering therefore different capsomer or environments (with two-fold, three-fold or fivefold symmetry) on its highest altitude as probed by the AFM tip. Yet, the effect was observed in the presence of viral DNA. Analysis of the AFM images allowed to correlate mechanical properties with capsid orientation.
In the case of bacteriophages, the difference in stiffness was traced back again to the orientation of the capsid during adhesion, but this was interpreted as a signature of the \emph{asphericity} of the virus rather than inhomogeneity at the single capsomer level.
However these observations of stiffness heterogeneity are not systematic, and for example it is not possible to detect significant variations of stiffness for some viruses like Cowpea Chlorotic Mottle Virus (CCMV) depending on its orientation \cite{michel2006}. As a consequence, the presence or absence of stiffness heterogeneity and its origin are far from being correctly understood. The purpose of the present work is to investigate quantitatively, both experimentally and theoretically, signatures for these mechanical heterogeneities. 

This paper is organized as follows. In the next section, we present our experimental approach based on AFM-nanoindentation for two different viral capsids, Adeno-Associated Virus 8 (AAV8) and Hepatitis B Virus (HBV, with T=4 capsid conformation), and then the numerical and analytical approach in order to predict and interpret  the results of indentation experiments quantitatively. Both approaches are combined and discussed in the Results section. These findings are then discussed in the light of past literature. Consequences of future experiments are also expressed.

\section{Experiments of AFM nanoindentation on viral capsids}
\subsection{AFM nanoindentation experiments}
\emph{AAV purification}

\begin{figure*}[ht]
\centering
\includegraphics[width=15cm]{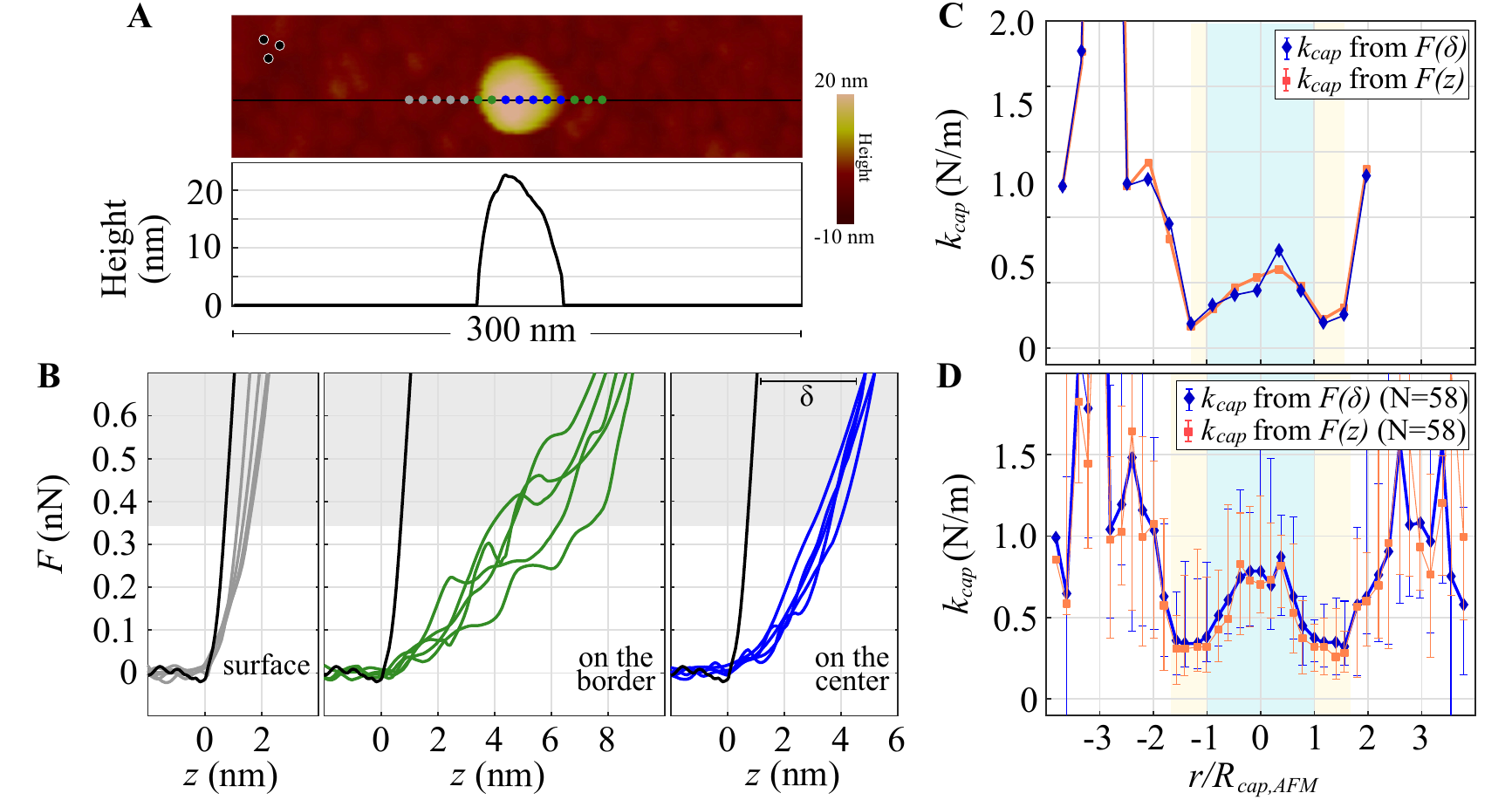}
\caption{\label{AFMFig1} Nanoindentation on AAV8 viral particles. \textbf{(A)} Typical image of a AAV8 viral particle deposited on surface. Indentations are performed at different locations represented by colored circles. Three additional indentations, shown by three black dots, were made away from the capsid to obtain a reference surface curve that contains only the cantilever deformation. The topographic profile along the black line is also shown. \textbf{(B)} Force-displacement curves for the different locations: \textit{(left)} grey points for virus-free locations, \textit{(center)} green points on the border of viruses, \textit{(right)} blue points in the center. The black curve represents an indentation made on the surface and away from the position of the virus (the best of the 3 \textit{surface} curves recorded away from the viral capsid. By subtracting at a fixed force the two z-positions on the black curve and a curve obtained within the capsid, it is possible to extract the capsid deformation $\delta$. \textbf{(C)} Single particle stiffness extracted at different locations, either through force-displacement $F(z)$ or force-indentation $F(\delta)$ curves. The indentation positions $r$ along the $x$-axis of the AFM image have been normalized using the capsid radius $R_{cap}$, AFM obtained from the height profile in (A) \textbf{(D)} Ensemble averaged stiffness at different locations ($N=58$ viruses). For each lateral position, the symbol (square or losange) is the median of the local stiffness distribution, and the bar represents the position of the first and third quartiles for the same distribution. }
\end{figure*}

Stock of AAV8 vector particles was produced and purified on CsCl gradients by calcium phosphate transfection of HEK-293 cells as described previously \cite{salvetti1998}. The vector plasmid AAVCMVeGFP, included the enhanced green fluorescence cDNA under the control of the cytomegalovirus promoter. The total length of the vector was 3769 bases, including the AAV inverted terminal repeats. The helper plasmids used for production were pDG8 (a kind gift from P. Moullier, INSERM U649, France). After purification and titration by qPCR,  the viral stock was kept at $-80^{\circ}C$.

\emph{HBV purification}

HBV capsids were obtained from the assembly of Cp183 proteins produced in bacteria following a protocol detailed in \cite{lecoq2018}. Cp183 capsids were expressed in \emph{E. coli} and purified similarly as described previously for truncated Cp149. Plasmid of pRSFT7-HBc183opt was transformed into E.coli BL21* CodonPlus (DE3) cells and grown at 37 $^\circ$C in LB medium culture. When OD600 reached around 2, expression was induced with 1 mM IPTG overnight at 20 $^\circ$C . Cells were collected by centrifugation at 6,000 g for 20 min. For one liter of culture, cells were resuspended in 15 ml of TN300 buffer (50 mM Tris, 300 mM NaCl, 2.5 mM EDTA, 5 mM DTT, pH 7.5) and incubated on ice for 45 min with 1 mg/ml of chicken lysozyme, 1X protease inhibitor cocktail solution and 0.5 $\%$ Triton X-100. 6 $\mu l$ of benzonase nuclease were added to digest nucleic acids for 30 min at room temperature. Cells were broken by sonication and centrifuged at 8,000 g for 1 h to remove cell debris. The supernatant was loaded onto a 10 to 60 $\%$ sucrose gradient buffered with 50 mM Tris pH 7.5, 300 mM NaCl, 5 mM DTT and centrifuged in SW-32Ti Beckman Coulter swinging bucket rotor at 140,000 g for 3 h at 4 $^\circ$C. Capsids were identified in gradient fractions by 15 $\%$ SDS-Polyacrylamide gel and precipitated by 40 $\%$ saturated $(NH_4)_2SO_4$. After incubation on ice for 1 h and centrifugation at 20,000 g for 1 h, pellets were resuspended in $10ml$ of purification buffer (50 mM Tris pH 7.5, 5 $\%$ sucrose, 5 mM DTT, 1 mM EDTA). The protein solution was centrifuged again for 15 min to remove insoluble pellet. The supernatant containing soluble capsids was dialyzed overnight against purification buffer at 4 $^\circ$C.  The purified CP183 capsids obtained in such an assembly process contain random short E Coli nucleic acids \cite{porterfield2010,heger2018} .

\emph{AFM sample preparation for AAV8 and HBV}
The viral stock of AAV8 or HBV capsids was diluted to a concentration of approximately $6x10^8$ capsids$/\mu l$ in TN $1mM$ (Tris $10\,  mM$ ($pH = 7.4$) and Nickel(II) chloride ($NiCl_2$) $1 mM$). Immediately after, $5 \mu l$ of the solution was deposited onto mica disks previously cleaned using adhesive tape. After 20 minutes of incubation to favor adhesion, we added $30 \mu l$ of TN $1mM$ to the sample and another $70 \mu l$ of TN $1mM$ in the AFM liquid imaging cell before imaging.

\subsection{AFM imaging and nanoindentation protocol}

The samples were imaged using a Bruker Nanoscope V Multimode 8 AFM using PeakForce Mode in liquid. We used three different cantilevers, SNL and ScanAsyst-fluid+ with a nominal tip radius of 2 nm and ScanAsyst-fluid with a nominal tip radius of 20 nm. The SNL and the two ScanAsyst cantilevers have a nominal spring constant of 0.35 and 0.7 N/m, respectively. We use mainly ScanAsyst fluid+cantilevers to compare both AAV8 and HBV nanoindentation and AFM imaging. We used the two cantilevers with a tip radius of $2nm$ (ScanAsyst fluid+) and $20nm$ (ScanAsyst fluid) to study the size tip effect in the AAV8 particles (supplementary figure 2). Also, we studied the effect  of the stiffness of the cantilever in AAV8 nanoindentation; for this,  we used the SNL and ScanAsyst fluid + cantilevers ( supplementary figure 3). 
Large scale AFM square images of 3 $\mu$m side (512*512 pixels) at 2 Hz scan rate were obtained to verify the deposition conditions on a large number of capsids. Figure 1 of supplemental material shows a zoom of those large AFM images for AAV8 and HBV capsids. 

The capsids were then imaged one by one to perform nanoindentation. Images were scanned at 2 $Hz$ over scan segments of 300-500 nm broad. The maximal force varied between 200-300 $pN$. Using Point and Shoot feature of Bruker acquisition software, we defined a line with 15 equidistant points over the capsid and 3 points far from the capsid to get the surface reference force curve (see Figure \ref{AFMFig1}A). Additionally, a horizontal height profile that passes through the highest point of the capsid is plotted and will be used to determine the capsid radius $R_{cap,AFM}$ used to normalize the stiffness profile before alignment and averaging. Nanoindentation data have been analyzed semi-automatically using a homemade Matlab code. Figure \ref{AFMFig1}B shows the force-displacement $F(z)$ curves separately according to the indentation position over the viral capsid. The black line is the best curve among the three recorded over the surface far from the capsid, and its slope represents the cantilever spring constant $k_c$. All those curves were aligned with the contact point.

\emph{AFM Nanoindentation data analysis }
We used two methods to extract the stiffness from the capsid.  
In the first method, we considered the force curve as a function of the cantilever displacement $F(z)$ that contains the deformation of both the cantilever and the capsid in the first linear regime. Assuming the capsid and the cantilever behave like two springs in series, the capsid stiffness $k_{cap}$ was calculated using the relation $k_{cap}=k(F(z)) = (k_{eff}k_{c})/(k_{eff}-k_{c})$. The cantilever stiffness $k_c$ is determined using the thermal noise method, and the effective stiffness $k_{eff}$ is the slope of the linear part of the $F(z)$ curve. 
In the second method, $k_{cap}$ is extracted from the Force indentation $F(\delta)$ curve as simply its slope. To obtain a $F(\delta)$ curve, we used the fact that for a given applied force $F$, the indentation $\delta$ is found by subtracting the deformation of the cantilever from the total tip displacement. Figure \ref{AFMFig1}C show the stiffness profile using both calculation methods. To superimpose N=58 differents indented capsid profiles, we aligned them according to the two lowest points of the profile. Then, the indentation positions $r$ were normalized using the capsid radius $R_{cap,AFM}$ obtained from the capsid height profile (Fig. \ref{AFMFig1} A). Finally,
we determined a unique stiffness profile where the diamond and square represent the median of all the $k_{cap}$ values for position zero, for example. The upper and lower error bars are 75 and 25$\%$ of the data (see figure \ref{AFMFig1}D).

\section{Modeling AFM nanoindentation on viral capsids}
\subsection{Thin shell model for intrinsic stiffness heterogeneity}
Thin shell elasticity is very well documented both in standard textbooks \cite{landau,ventsel,calladine,timoshenko} and outstanding research publications \cite{seung1988,bowick2000,travesset2005}. We will mainly recall in this subsection the main results of interest for our study. In particular, we present scaling derivations for the stiffness, which have been shown to be appropriate in the regime of linear elasticity. Within standard 3D elasticity, the deformation of a material from an unconstrained configuration is characterized mainly by two tensors: the strain tensor which quantifies the relative material deformation, and the stress tensor which quantifies the strength and orientation of forces arising from this deformation. In the simplest case of unidimensional material, the linear relationship between longitudinal strain $\epsilon$ and stress $\sigma$ is called Hooke's law and it defines the Young modulus as $Y=\sigma/\epsilon$, which is an elementary measurement of the stiffness of the material. 

The concept of stiffness can be extended to different material geometry. In the case of thin shells like viral capsids, it is defined as the ratio between the applied mechanical force and the indentation length. In the case of a homogeneous spherical thin shell of width $t$ and radius $R$, the stiffness results from the balance of in-plane stretching/compression and out-of-plane bending. Introducing the 2D stretching modulus as $\kappa_s=Yt$ and the 2D bending modulus (also known  as the flexural rigidity) as $\kappa_b=Yt^3/(12(1-\sigma^2))$ where $\sigma$ is the Poisson ratio of the material, we define the dimensionless number associated to stretching/bending balance, which is also known as the F\"{o}ppl-von Karman number (FvK):
\begin{equation}
\gamma =\frac{\kappa_s R^2}{\kappa_b}
\end{equation}
The scaling of stiffness is obtained by estimating the energetic budget associated with small indentation $\Delta$ of the spherical shell \cite{landau,buenemann2008,komura2005,widom2007}. This indentation modifies  the curvature of the shell locally over a distance $d$ from a value $R^{-1}$ to $\rho^{-1}\sim R^{-1}-2\Delta/d^2$. The bending cost is of order $\Delta E_b \sim \kappa_b (\rho^{-1}-R^{-1})^2d^2\sim \kappa_b \Delta^2 /d^2$. The meridians are also compressed by the indentation resulting in a strain $\epsilon\sim \Delta/R$. The compression cost is of order $\Delta E_s\sim\kappa_s\epsilon^2d^2\sim\kappa_s d^2\Delta^2/R^2$. The actual deformation spread $d$ is obtained by minimizing the sum of bending and compression energies leading to $d_e\sim R^{1/2} (\kappa_b/\kappa_s)^{1/4}\sim R\gamma^{-1/4}$. The net energy associated with the indentation is therefore written as $\Delta E_{tot}\sim k \Delta^2$ with the stiffness
\begin{equation}
\label{k landau}k\sim \left(\frac{\kappa_s\kappa_b}{R^2}\right)^{1/2}\sim\kappa_s\gamma^{-1/2}
\end{equation}

This result is strictly valid for small enough indentation, and more importantly for small FvK numbers. Indeed, it has been shown by Lidmar \textit{et al.} that closed shells with large FvK numbers are strongly \emph{faceted} with icosahedral symmetry \cite{lidmar2003}. This phenomenon is related to the presence of topological defects, not taken into account in the scaling derivation presented so far. From a structural point of view, these defects are associated with a few numbers of capsomers (local group of proteins) having five proteins, while all other capsomers are composed of six proteins. This can be understood first from a geometrical point of view: the virus shape and in particular its icosahedral symmetry can be reproduced by a simple triangulation network model in which each triangle represents three proteins, according to the classical Caspar and Klug approach \cite{caspar1962}. Now equilateral triangles in a hexagonal phase are known to produce  a perfect tiling of a flat plane or any surface having zero curvature in at least one direction. In contrast, pentagons made of equilateral triangles form  a 3D conical-like structure naturally and therefore curve  a surface locally in two distinct directions. This observation can be made more precise by considering the distribution of stress within thin shells. The equations describing the mechanical equilibrium of thin shells are known as the two F\"{o}ppl-von Karman equations. These highly non-linear coupled equations involve the out-of-plane deflection $w(\mathbf{r})$ and the in-plane stress $\sigma_{ij}(\mathbf{r})$. More precisely, with the introduction of Airy stress function $\chi$, which is related to the stress tensor $\sigma_{ij}$ by $\sigma_{ij}=\epsilon_{ik}\epsilon_{jl}\partial_k\partial_l\chi$, where $\epsilon_{ij}$ is the antisymmetric unit tensor, the second FvK equation reads
\begin{equation}
\frac{\nabla^4\chi(\mathbf{r})}{\kappa_s}=s(\mathbf{r})-K_G (\mathbf{r})
\end{equation}
where $s(\mathbf{r})$ is the defect density and $K_G(\mathbf{r})$ is the Gaussian curvature \cite{seung1988}. With this equation, it is straightforward to realize that the morphology of a surface with non-zero Gaussian curvature is necessarily coupled to the presence of defects, in order to minimize the elastic stress within the shell. In the simplest case, these defects known as \emph{disclinations} are the pentamers. Moreover, the number of these defects is fixed by the topology of the surface, as it can be seen by Euler relation \cite{giomi2007}. For closed shells, this constraint imposes to have exactly twelve pentamers. It has been shown by several authors that most structures obtained using self-assembly will have defects regularly spaced, giving rise to the icosahedral symmetry previously mentioned \cite{li2018,menou2019}. 

As the presence of these defects is obviously related to the mechanical properties of the shell, it is expected that the stiffness is different at these locations.  Several authors have analyzed both numerically and experimentally the influence of these defects \cite{buenemann2008,widom2007,gibbons2008,klug2012,llauro2016}. In the limit of small Fvk numbers, for which the shell shape is spherical, it is possible to extend the previous scaling for the stiffness estimate by modifying the stretching contribution to the energy. Indeed, for a hemispherical cap of curvature $R^{-1}$ and lateral extension $d$, subjected to vanishing boundary force $\sigma_{rr}(r=d)=0$ \footnote{This approximation has been made for the sake of simplicity. We checked that the scaling results are not altered when non-zero boundary force $\sigma_b$ is taken into account}, the in-plane stretching energy yields $\Delta E_{s}= (\pi/384) \kappa_s  d^6/R^4$ \cite{schneider2005,grason2012,azadi2012} . Note that this formula allows recovering previous scaling result for stretching energy as function of indentation $\Delta$ by considering hemisphere's curvature change $R^{-1}\rightarrow\rho^{-1}\sim R^{-1}-2\Delta/d^2$. In the presence of a central disclination of charge $s$, the stretching energy becomes 
\begin{equation}
\Delta E_{s5}=\Delta E_s+\frac{\kappa_sd^2}{32 \pi}\left(s^2-\frac{\pi s d^2}{R^2}\right)
\end{equation}   
Expanding this energy in $\Delta$ upon curvature change, leads to correct the stiffness associated with indentation on pentamers ($s=\pi/3$) to 
\begin{equation}
k_5\sim k_6-\frac{\pi\kappa_s}{48}
\end{equation}
with $k_6\sim\kappa_s \gamma^{-1/2}$.  
This estimate shows that there is a slight reduction in stiffness associated with pentamers but its amplitude is not expected to be significant.

In the opposite limit of large FvK number, the scaling behavior has been derived by Widom \textit{et al.} \cite{widom2007}. In this limit, the viral capsid becomes faceted with icosahedral symmetry, and indentation experiments are expected to give different results, whether a vertex (\textit{i.e.} a pentamer) or a face (\textit{i.e.} a hexamer) of the icosahedron is probed. In the latter case, indentation is performed on a flat triangular face. The resulting stiffness in this idealized geometry can be found in the classical textbook by Timoshenko \cite{timoshenko} by solving the first FvK equation for the out-of-plane deformation profile with appropriate boundary conditions. The result scales like 
\begin{equation}
k_{face}\sim\frac{\kappa_b}{R^2}\sim\kappa_s \gamma^{-1}
\end{equation} 
Finally, upon indentation of an icosahedron's vertex, the force is mostly transmitted within the plane of each face. This problem has also been  considered in the classical textbook by Landau \cite{landau}. In this case, the in-plane deformation induced by a force localized at the edge of a face has been shown to depend in a logarithmic way on the distance to the edge. The stiffness scales therefore as:
\begin{equation}
k_{vertex}\sim \frac{\kappa_s}{\ln R/a} \sim \kappa_s (\ln \gamma)^{-1}
\end{equation}
where $a$ is a small cutoff distance associated with the presence of the core of the pentamer. The comparison of these two stiffnesses $k_{face}$ and $k_{vertex}$ is expected to be more favorable to experimental discrimination. In particular, the stiffness of pentamers is expected to be larger in this regime. These results have been derived by Widom \textit{et al.} using the expected elastic behavior of an icosahedron, so they should be valid for whatever shell structure that fulfills this symmetry. This is expected to be the case for large FvK numbers, where the dominant stretching cost over bending induce such a shape, but it might also be valid for very small structures which are intrinsically faceted like $T=1$ viral capsid. These structures are so small that they are composed of pentamers only, giving rise to a strongly faceted shape. 

\subsection{Numerical simulations of nanoindentation}
\begin{figure*}[ht]
\centering
\includegraphics[width=15cm]{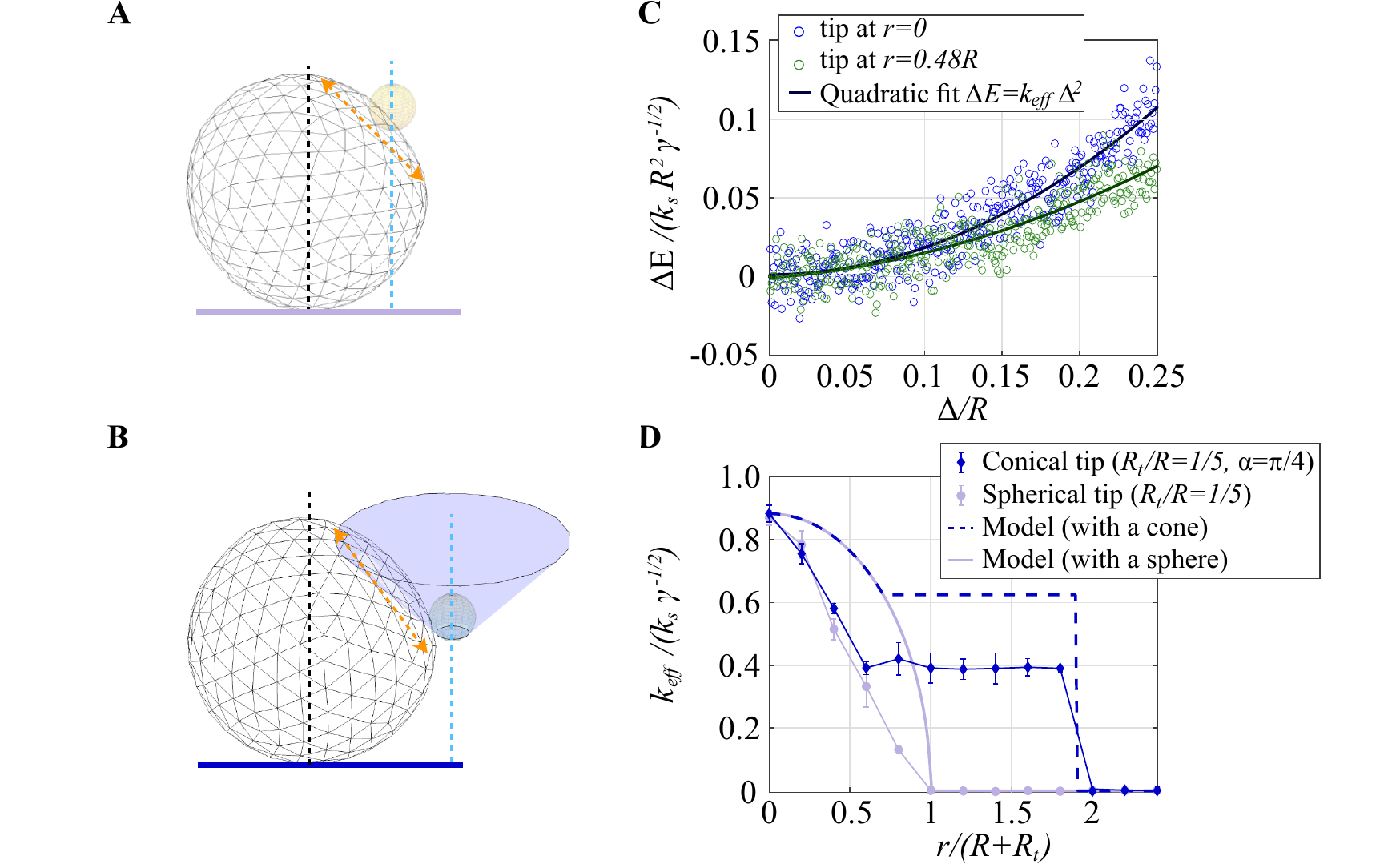}
\caption{\label{Numerics} Simulation of nanoindentation experiments. Snapshots of the indentation process of the triangulated surface by a spherical tip (\textbf{A}) or a conical tip (\textbf{B}). Orange double-arrows shows the area in which  the deformation is not negligible. (\textbf{C}) Elastic energy as a function of the relative indentation $\Delta/R$ for two positions. Quadratic fit and parameters are also shown. (\textbf{D}) Effective stiffness as a function of the lateral position of the tip for sphere and cone. Simple geometrical model (thin and dashed lines) for  effective stiffness. The stiffness is renormalized with classical prediction Eq.\ref{k landau}.
}
\end{figure*}
In order to complement our analytical and experimental approaches, we developed numerical simulations of the indentation of thin shells. The elastic properties are reproduced by using the classical model of a triangulated surface. Within such a numerical model, in-plane deformations are taken into account with bond deformations, varying their lengths, whereas out-of-plane deformations are taken into account varying dihedral angles between adjacent faces or triangles \cite{seung1988}. The elastic energy is written as 
\begin{equation}
E_{elas} = \sum_{\langle i,j \rangle} \frac{1}{2}k_s (\ell_{\langle i,j \rangle} - \ell_0)^2 + \sum_{\alpha,\beta} k_b(1-\cos \theta_{\alpha,\beta} ), 
\label{F_d}
\end{equation} 
where $\ell_{\langle i, j \rangle}$ is the length of the edge $\langle i, j \rangle$. When the latter length is different from the preferred one $\ell_0$, the bond is stretched or compressed by a spring force proportional to the spring stiffness $k_s$. Furthermore, $k_b$ sets the bending energetic cost for non-zero angle $\theta_{\alpha, \beta}$ between the two given adjacent faces $\alpha, \beta$. Following the works of Seung \textit{et al.} for example, the relation between stretching and bending modulus in the analytical and numerical model is given by \cite{seung1988}
\begin{equation}
k_s=\frac{\sqrt{3}}{2}\kappa_s \, ,\, k_b=\frac{2}{\sqrt{3}}\kappa_b
\end{equation}
In order to mimic as much as possible real viral capsids, we analyze the structure of AAV8, which has a $T$ number of $1$. Since this small structure might be not large enough in terms of a triangular subunit, we divide each capsomer with a  $5 \times 5$ subtriangular lattice preserving the icosahedral symmetry and the non-skew capsid shape. For AAV8, we used a radius of $R=12.5\,nm$.

The thin shell interacts both with the flat substrate on which it is lying, and with the approaching spherical or conical tip. This is realized by adding an adhesion energy to all vertices $E_{ads}=\sum_{i}V(\mathbf{r}_i)$. For both interactions, we chose a Morse potential $V(r) = V_0(1 - e^{-r/a})^2$, which is naturally repulsive at a very short distance, and attractive at a larger distance. In the limit of very large distance, the adhesion force vanishes exponentially. We added a steric cost to each face to prevent the tip from entering the triangular network, \emph{i.e} we also consider  the repulsion of the barycenter of each face. Contact points  with the substrate modeled as an infinite rigid flat plate are not allowed to slide tangentially and fixed at their initial position. In our simulation, we set $V_0 = 2k_B T \simeq 1.2\, kcal/mol$ and $a = 2\, \AA$. For some simulations, we forbade the motion of bottom vertices, in order to prevent the shell from rolling.

The relaxation is done using  Langevin dynamics:
\begin{equation}
\mathbf{r}_i(t+dt) = \mathbf{r}_i(t) - \frac{dt}{\zeta}\mathbf{\nabla} (E_{elas} + E_{ads}) + \left(2 \frac{k_B T}{\zeta} dt \right)^{1/2} \mathbf{\eta},  
\end{equation}
where $\mathbf{r}_i$ represents the position of the vertex, $\eta$ a random force and $\zeta$ the damping coefficient. We chose $\zeta = 1 kcal/mol/nm$ to get good agreement with the experimental diffusion time of a capsomer, and the integration time-step $dt = 400\, fs$. We assumed quasi-static nanoindentation, so that the tip is lowered by $9 \,pm$ between each relaxation iteration that lasts $2\, \mu s$. We ensured that an equilibrium state is reached between each tip descent. Finally, since we restricted our investigations on spherical viruses or weakly faceted viruses for which the FvK number $\gamma \simeq O(1)$ or 10, we set $k_s = 200 k_B T \simeq 120 kcal/mol$ and $k_b = 80 \,k_B T \simeq 50 \,kcal.mol^{-1}$ at ambient temperature $T = 298K$.

From the simulations, the elastic energy of relaxed configurations are obtained, the position of the center of the AFM tip is saved and substracted with its initial position giving the indentation depth. The maximal indentation depth is set to $\delta_{\text{max}} = 0.3 R$, provided the tip has not reached the substrate in the off-centered configurations. When coordinates of the tip in the $x-y$ plane are not zeros, the position of the tip is carefully computed so that indentations begin at the contact of the numerical cap. Data are taken between each relaxation step, and each nanoindentation experiments is repeated several times. A quadratic fit of the form $E_{elas}=\frac{k}{2}\Delta^2$ using the \texttt{Nonlinearfit} method of \texttt{Mathematica} enables us to extract the effective stiffness $k$ of the virus for each experiment, and error bars are the corresponding root mean-square deviations resulting from the repetitions.

\subsection{Geometrical model for position-dependent stiffness}
\begin{figure*}[ht]
\centering
\includegraphics[width=20cm]{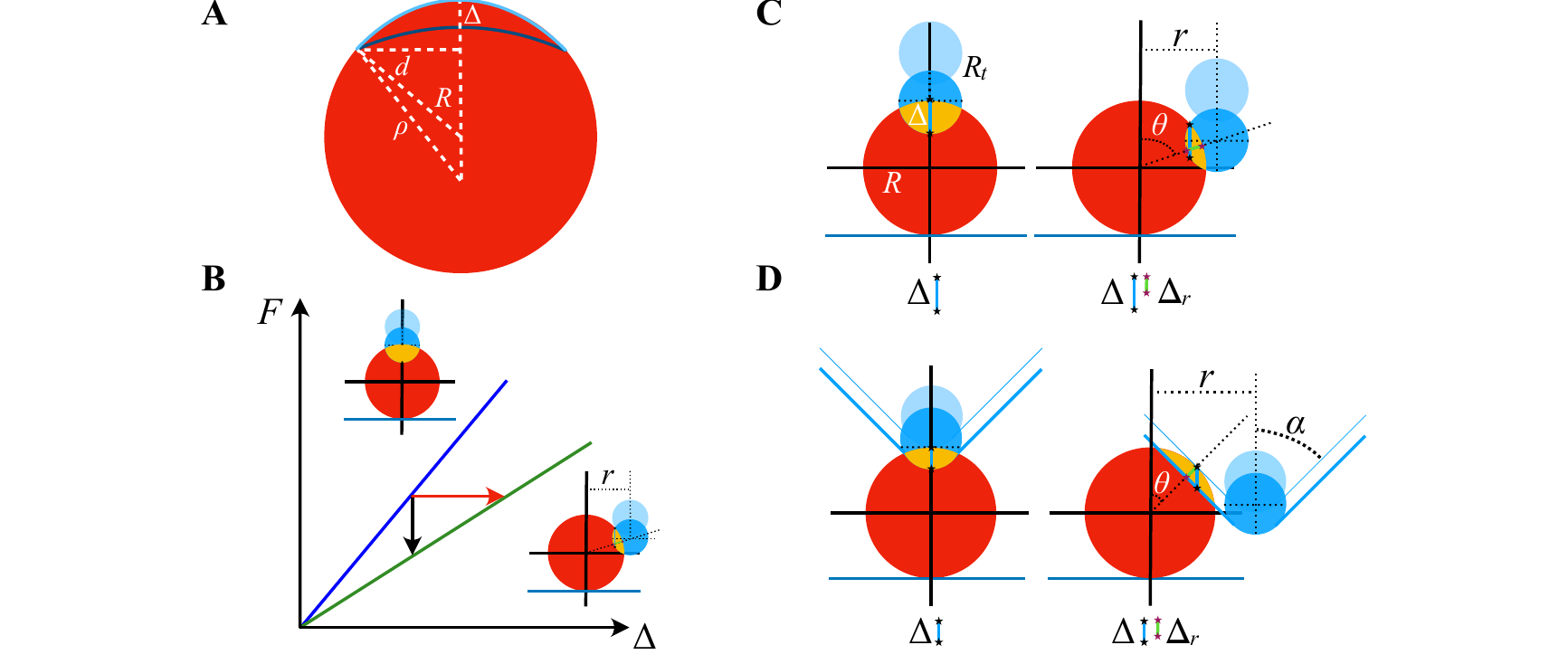}
\caption{\label{GeometryParameters} Geometrical model for indentation. (\textbf{A}) The shell to be indented has a radius $R$. Upon vertically indenting on a length $\Delta$, the shell surface deforms horizontally over a typical length $2d$. The deformed area is assumed to have a constant curvature radius $\rho$. (\textbf{B}) Ideal force-indentation curve in the linear regime. Vertical downward arrow represents the change of force needed to perform similar vertical indentation at a different lateral position. Horizontal rightward arrow represents the change in indentation reached at a similar force at a different lateral position. (\textbf{C}) and (\textbf{D}) True deformations (yellow areas) achieved by identical vertical indentation for two different lateral positions, for a spherical tip \emph{(C)} or conical tip  \emph{(D)}}.
\end{figure*}
The scaling results from the previous section assume that the indentation is performed thanks to a point-like force applied precisely on the top of the virus. A real AFM tip has, of course, a finite radius of curvature of order few tens of nanometers, and its value is often similar to the radius of curvature of the virus itself. This observation suggests that great care should be the rule while considering the geometry of indentation experiments. Anticipating any misalignment of the tip with the highest point of the virus, we model in this section the variation of stiffness while several lateral positions relative to the top point of the shell are vertically indented with a finite size tip. We consider two types of tip: a spherical tip and a conical tip ending with a hemisphere.

The precise geometry considered, and the definition of the model parameters, are shown in figure \ref{GeometryParameters}. In the case of a spherical tip, these cartoons illustrate an obvious effect that leads to strong variation of measured stiffness as the tip goes from the top to the edge of the spherical shell: upon performing identical \textit{vertical} indentations at different lateral positions from top to the edge, the corresponding shell deformation decreases while approaching the edge. In the figure, this is highlighted by simply observing the intersection of the two spheres. However, real indentation is expected to induce deformation at a larger scale, as it was already discussed in the previous section. Accordingly, the spreading of the deformation is estimated by $d_e\sim R\gamma^{-1/4}$. Nevertheless, the estimation of the overlap region provides a simple way of quantitatively evaluating the influence of the lateral tip position. In particular, the quantity of interest is the \textit{radial} indentation length $\Delta_r$, which represents the maximal radial deformation along the new rotated axis in direction $\theta$. After a lengthy but straightforward geometrical calculation, one gets for a thin shell of radius $R$ and a spherical tip of radius $R_t$:
\begin{equation}
\Delta_{r}=\frac{1+\frac{\tilde{\Delta}}{R-\tilde{\Delta}}-\sqrt{1+\left(\frac{\tilde{\Delta}}{R-\tilde{\Delta}}\right)^2\left(\frac{1-\cos\theta^2}{\cos\theta^2}\right)}}{1+\frac{\tilde{\Delta}}{R-\tilde{\Delta}}}
\end{equation}
together with the relation between lateral tip position $r$ and angle $\cos \theta =\sqrt{1-\left(\frac{r}{R+R_t}\right)^2}$ and the new variable $\tilde{\Delta}=\Delta\frac{\cos \theta}{1+\frac{R_t}{R}}$. In the limit of small indentation, this result simplifies to $\Delta_r\simeq\Delta\sqrt{1-\left(\frac{r}{R+R_t}\right)^2}$. A weaker deformation of the  shell in the radial direction is therefore associated with a weaker elastic response (see Fig.\ref{GeometryParameters}b). Under the assumption of linear elasticity, the effective stiffness which is position-dependent is written therefore as
\begin{equation}
\label{ksphere}k_{eff}^{(sphere)}(r)=k(0)\sqrt{1-\left(\frac{r}{R+R_t}\right)^2}
\end{equation}

In the case of a conical tip of opening angle $\alpha$, and terminated with a hemisphere of radius $R_t$, the \emph{radial} indentation also depends on the position, at least when the spherical part of the tip is in contact with the shell. However, when the conical part is in contact with the shell, the radial indentation associated with a vertical indentation $\Delta$ is identical whatever the lateral position. This is because one of two principal curvatures of the cone is zero, and therefore the overlap with spherical shell along this direction is the same whenever there is a contact. More precisely, the indentation is independent of the position for $r>(R+R_t)\cos\alpha$, \textit{i.e.} when $\theta=\pi/2-\alpha$, and its constant value is simply $\Delta_r=\Delta \sin\alpha$. The stiffness is therefore
\begin{equation}
\label{kcone}k_{eff}^{(cone)}(r>(R+R_t)\cos\alpha)=k(0)\sin \alpha
\end{equation}
Note that although the radial indentation is constant, it is expected that transverse deformation slightly increases with the position as the conical imprint on the shell becomes larger. The two estimations Eqs. \ref{ksphere} and \ref{kcone} are simply based on the geometry of shapes overlapping with the shell. It can be complemented, for example, by the estimation of the boundary length of the overlap region. This is demonstrated in the appendix. Nevertheless, the major position-dependent effect is found with the true indentation length.  
 
The previous estimation assumes that the shell does not move laterally upon vertical indentation due to strong adhesion with the substrate. In the limit of weak adhesion, the shell is indeed expected to escape the imposed constraint of indentation by rolling or sliding sideways. In the intermediate adhesion regime, the shell will be both deformed and slightly shifted sideways. This additional shift in the position is expected to lower the true indentationfurther, and therefore this effect might reduce further the effective stiffness. The relevance of this effect can be simply estimated the following way. We assume that the vertical indentation produces, in addition to the true indentation along the symmetry axis of the deformation, a horizontal shift $\Delta_h$. The relation between vertical indentation $\Delta$, true indentation $\Delta_r$ and horizontal shift is written as $\Delta_r=\Delta \sqrt{1-\left(\frac{r+\Delta_h}{R+R_t}\right)^2}$ to the lowest order in $\Delta$. The equilibrium is maintained by assuming the presence of a horizontal restoring force $|F_{adh}|=k_{ads}\Delta_h$, which balances the horizontal contribution of the deformation force $|F_{deform}|=k(0)\Delta_r\frac{r+\Delta_h}{R+R_t}$. To the lowest order, the horizontal shift is finally estimated as
\begin{equation}
\label{Deltah}\Delta_h\simeq\Delta\frac{k(0)}{k_{ads}}\frac{r}{R+R_t}
\end{equation}
Therefore the horizontal shift increases with lateral position, and so does the horizontal force associated with adhesion bonds. This estimate shows that adhesion bond rupture might become more likely as the edge of the particle is approached and that the shell might slip or roll away from the tip. Note that the horizontal shift decreases for larger adhesion strength, as it is inversely proportional to $k_{ads}$.

\section{Comparing theory and experiments of nanoindentation of viral capsids}
\subsection{Geometrical stiffness heterogeneity}
According to the protocol described in the previous experimental sections,  we performed nanoindentation experiments with AFM on two different viral capsids, AAV8 and HBV (T=4 capsids) (Fig.\ref{AFMFig1}). For each viral particle to be investigated, we recorded force-displacement curves at different lateral positions across the particle (Fig.\ref{AFMFig1}A). The resulting curves for one particular, but representative capsid, is shown in figure \ref{AFMFig1}B, in which all the curves have been horizontally aligned such that force raises at the same vanishing displacement.

These curves show clearly that the elastic response is strongly dependent on the lateral position of the tip with respect to the top of the particle. Using the average best force-displacement curve recorded for the substrate, one can obtain a similar picture on the force-indentation curves (Fig.\ref{AFMFig1}C). We recall here that the indentation $\delta$ is the net deformation of the shell with respect to its initial configuration prior to tip-shell contact. Both types of curve exhibit some oscillations when vertical indentation is performed far from the top of the shell (Fig.\ref{AFMFig1}B). This is likely to be associated with a partial loss of substrate adhesion: as it was discussed in the analytical model, 
vertical indentation performed at large distance from the top of the shell induces a weak deformation of the shell, but also exerts an increasing horizontal force on the adhesion bonds between the capsid and the substrate. 

Assuming a linear elastic response, the force curves can be used to estimate the stiffness, as it is shown in figure \ref{AFMFig1}C, where this information is represented as a function of lateral position for a single particle. This demonstrates very clearly that the effective stiffness decreases as the indentation experiment is performed on even not so distant lateral positions from the capsid's top. This observation survives when the results on several particles are aggregated (Fig.\ref{AFMFig1}D). Notice though that due to the oscillations in the force-displacement curves for large lateral positions, the effective stiffness in these regions has a larger uncertainty. 
We also performed  similar indentation measurements on HBV (T=4 viral capsids). In this case, we observed position-dependent stiffness, although with a weaker amplitude, as will be discussed later.

In order to investigate more quantitatively this position-dependent stiffness and its dependence on various parameters of the system, we used the molecular dynamics simulations described in the previous modeling sections. The indentation process is reproduced by computing the change in elastic energy of a triangulated surface constrained by the presence of a spherical or conical tip (fig.\ref{Numerics}A and B) and a flat substrate. 

Fitting the expected quadratic dependence of elastic energy under the assumption of linear elasticity, we extract the stiffness for different lateral positions (fig.\ref{Numerics}C). The resulting stiffness variation, shown in figure \ref{Numerics}D, is similar to the observed experimental stiffness change. Using numerical simulations also allows  to investigate the influence of various parameters on the position-dependent stiffness such as tip geometry (spherical or conical) and elastic parameters. 


Comparing both experimental and numerical indentations, we observed first a strong variation of the stiffness. This effect can be qualitatively understood by a simple geometrical model previously mentioned: a vertical indentation performed at different lateral positions will correspond to weaker effective deformation of the shell, resulting in a weaker restoring force. From a quantitative point of view, this geometrical model provides a simple prediction that can be compared to both experiments and simulations. In the case of a spherical tip, the expected stiffness decreases toward vanishing value, and the simulations confirm this prediction. On the contrary for a conical tip, the stiffness should first decrease similarly to the spherical case, and then reach a non-zero constant value. Again, this is also observed both in the experiments and in the simulations. However, the latter curves show a stronger decrease in the stiffness compared to the simple geometrical model, suggesting the presence of other effects at play in the real system. 
\begin{figure*}[ht]
\centering
\includegraphics[width=15cm]{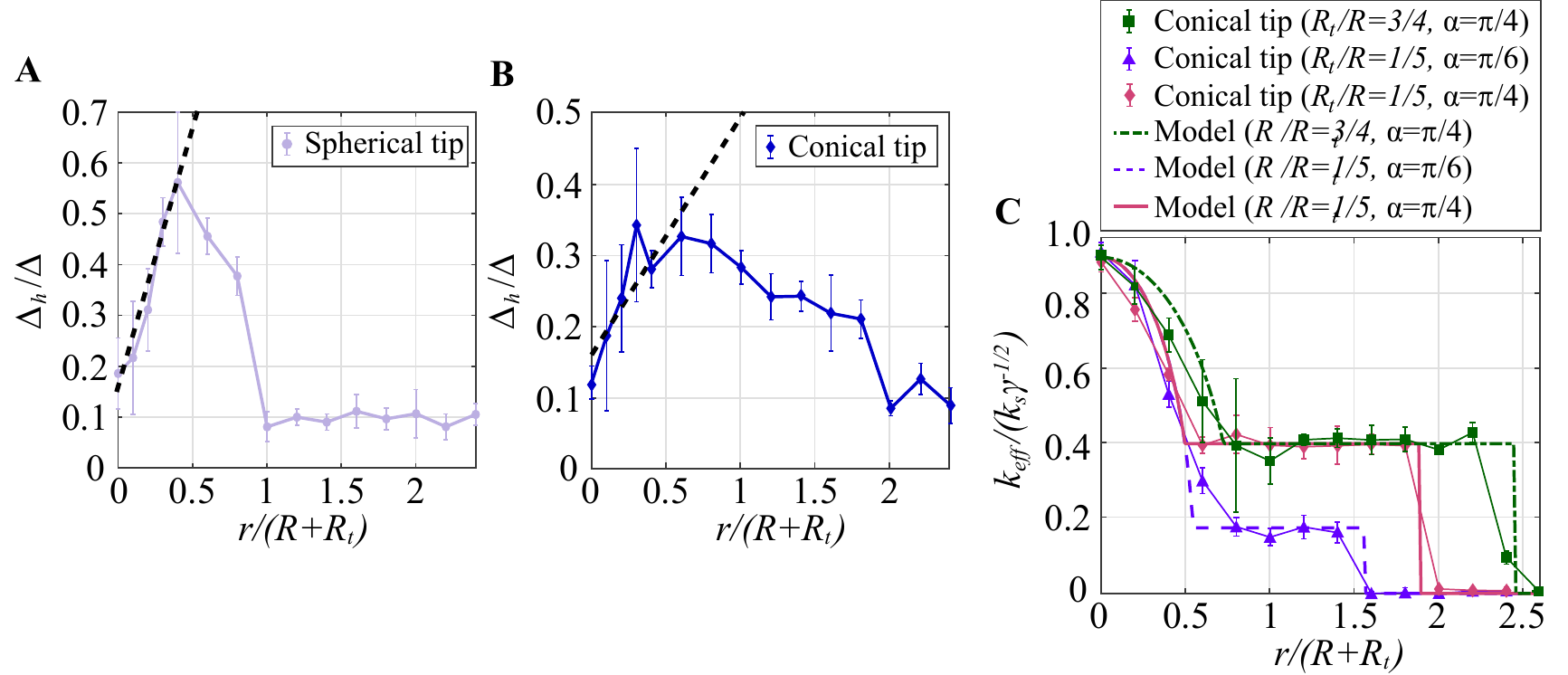}
\caption{\label{GeoParam}Lateral displacement upon off-centered vertical indentation in the simulation. The lateral displacement of the center of mass is shown as a function of the lateral position of the tip for (\textbf{A}) spherical or (\textbf{B}) a conical tip. The dotted lined shows the initial proportionality with the lateral position of the tip. (\textbf{C}) Stiffness profiles for various geometrical parameters, including the radius of the shell $R$, the radius of the tip $R_t$ and the angle of the cone $\alpha$. The corrected geometrical model taking into account the leading order for horizontal displacement is shown for various configurations with dashed lines.}
\end{figure*}

The first major correction that can be brought to the geometrical model is to consider horizontal displacement which is concomitant with the vertical indentation. Within the simulations, we can indeed compute for each lateral position of the tip, the net horizontal shift of the shell center of mass as compared to the original position prior to indentation. This is shown in figure Fig\ref{GeoParam} A and B,  respectively for spherical and conical tip. It is observed within these plots that the horizontal shift increases first linearly with the lateral position, in qualitative agreement with scaling model prediction Eq.\ref{Deltah}. This horizontal shift could be associated with partial loss of adhesion sites between the shell and substrate or with a more global deformation of the shell. Plugging a linear horizontal displacement $\Delta_h=Ar$ in the original geometrical model introduces a rescaling correction to the geometrical model Eq.\ref{ksphere} in the form $k_{eff}^{(sp)}=k(0)\sqrt{1-(r/(R+Rt))^2(1+A)^2}$. This predicts indeed a stronger decrease of stiffness as compared to the original geometrical model. This argument can also be applied to the conical model. Using $A$ as an adjustable parameter, the corrected geometrical shows a better agreement with experimental (supplementary figure 2) and simulated data (Fig. \ref{GeoParam}C).  

Simulations were also used to address the influence of various parameters on the effective stiffness measured. This is shown in figure \ref{GeoParam}C, for example. The geometrical model predicts that the value of constant stiffness for the conical tip does depend only on the opening angle of the cone (see Eq. \ref{kcone}) but not on the value of sphere radius terminating the cone. This behavior is observed in the figure with two identical plateaux of stiffness for two different sphere radii. Moreover, the value of the plateau is expected to increase with $\alpha$. Again, this prediction is quantitatively reproduced in the simulations (Fig. \ref{GeoParam}C).
 
\subsection{Intrinsic stiffness heterogeneity} 
Up to this point, we demonstrated and discussed stiffness heterogeneity depending strongly on the lateral position of the tip. This raises the question of monitoring intrinsic stiffness heterogeneity due to the shell's structure itself. In order to address this question, we changed the value of elastic parameters and designed specific numerical experiments. The most promising candidates for intrinsic stiffness heterogeneity are the pentamers and hexamers of the shell. We imposed the orientation of the shell on the substrate such that either a pentamer or a hexamer is facing up the tip. For this last case, there are several possible choices. However, the simplest case is to choose a hexamer equidistant from three neighboring pentamers, \textit{i.e.} this corresponds to any 3-fold symmetry axis of the icosahedron. In other words, this hexamer is supposed to lie at the center of the faces in the faceted shell representation. Next we induced the same vertical indentation for different lateral positions, and we deduced the effective position-dependent stiffness.
\begin{figure*}[ht]
\centering
\includegraphics[width=15cm]{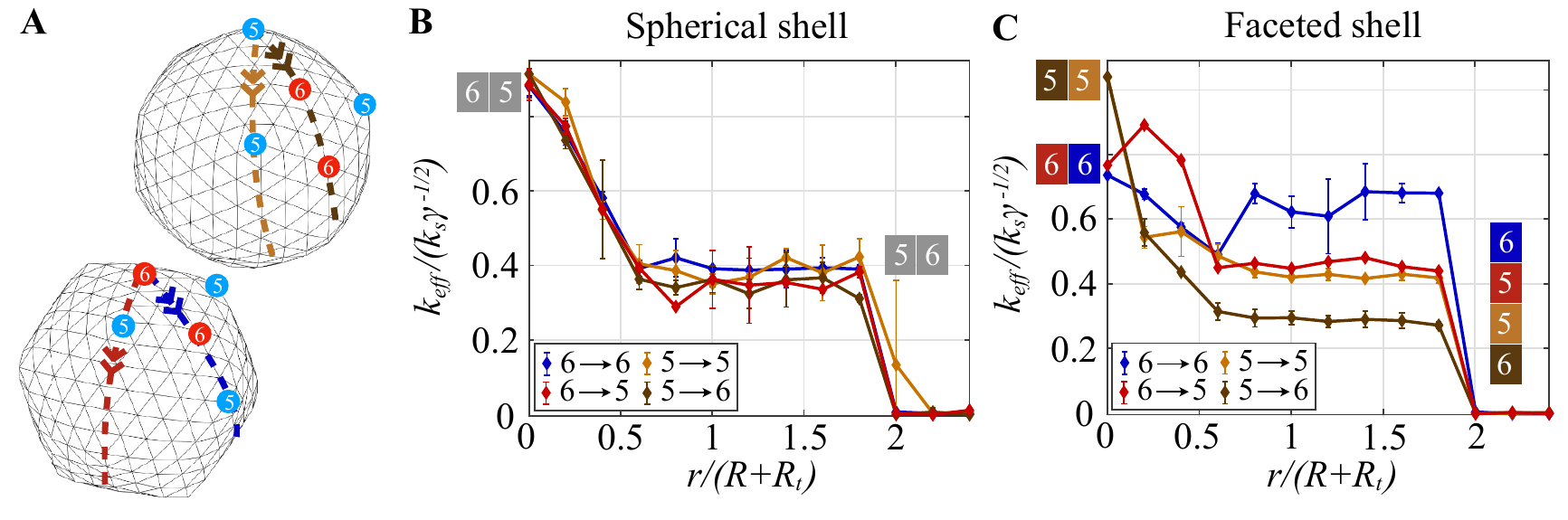}
\caption{\label{FvKInfluence}Heterogeneity of stiffness profiles in the simulation. The profiles corresponding to particular paths along the shell's surface are shown in (\textbf{A}) as colored dashed lines on faceted or spherical shells. The stiffness values associated with the profiles are shown for faceted shells with $\gamma\simeq 576.5$ (\textbf{B}) or spherical shell with $\gamma\simeq 57.65$ (\textbf{C}). The underlying connectivity of the shell was kept constant and correspond formally to the $T=9$ icosahedral shell. Four trajectories have been considered in both cases: pentamer to pentamer \textit{(bright maroon)}, pentamer to hexamer \textit{(dark maroon)}, hexamer to hexamer \textit{(blue)} and hexamer to pentamer \textit{(red)}. Profiles are similar for spherical shells, while they show some heterogeneity for faceted shells.}
\end{figure*}
For the sake of simplicity, we used $T=9$ shells to performed nanoindentation experiments along different lateral profiles, and we achieved different degree of faceting by using different elastic parameters.  Using this protocol, we were able to compare the stiffness for different \textit{tip trajectories}: from central pentamer to peripheric pentamer, central pentamer to peripheric hexamer, central hexamer to peripheric hexamer, central hexamer to peripheric pentamer. Note that this procedure is different from most numerical works in the literature, where indentation is performed on top of the shell for the various shell orientations encountered by the adsorption process on the substrate. 
At low FvK number or equivalently for roughly spherical shells, we observed that the stiffness decrease as a function of the position is similar, whatever the trajectory (Fig.\ref{FvKInfluence}B). This agrees with the scaling prediction of thin shell theory as presented earlier. In this regime, the geometrical stiffness heterogeneity is dominant, and intrinsic heterogeneity is likely to be very hard to measure in real AFM data, as suggested by the observation in the simulations.

On the contrary, we anticipate different behavior for conditions where faceted shells are expected, namely at higher FvK number or for $T=1$ shells. In this case, the stiffness variation upon different tip trajectories are distinct (Fig.\ref{FvKInfluence}C). First, the pentamers on top of the shell are stiffer than hexamers. This is also confirmed by the value of the stiffness plateau: for trajectories starting from a central pentamer, the plateau is higher if the trajectory ends on a peripheric pentamer than on a peripheric hexamer. The analysis of the last two trajectories is more involved, as it results from the balance of both intrinsic and position-dependent stiffness. Going from one central hexamer toward a distal pentamer, the stiffness first increases, reflecting the intrinsic or structural variation, before decreasing at larger distances, showing the geometrical effect. This could be due to the presence of a ridge joining two pentamers on the trajectory. Finally, going from one central hexamer to a peripheric hexamer might end up in the deformation of aligned pentamers, for a large lateral position. The dispersion of trajectories shows nevertheless that the effective position-dependent stiffness is strongly dependent on the trajectory itself. At this stage, we are led to the following partial conclusion: a closed shell that has a roughly spherical shape does not exhibit significant intrinsic stiffness heterogeneity, as all stiffness profiles superimpose. Therefore the presence of pentamers and hexamers does not lead necessarily to stiffness heterogeneity. On the contrary, when the shell exhibits significant faceting, stiffness heterogeneity is observed, and different trajectories lead to different stiffness profiles.

Next, we tried to observe similar heterogeneity signatures in the experiments on both viral capsids AAV8 $(T=1)$ and HBV $(T=4)$ using several indentations performed at various lateral positions. 
\begin{figure*}[ht]
\centering
\includegraphics[width=15cm]{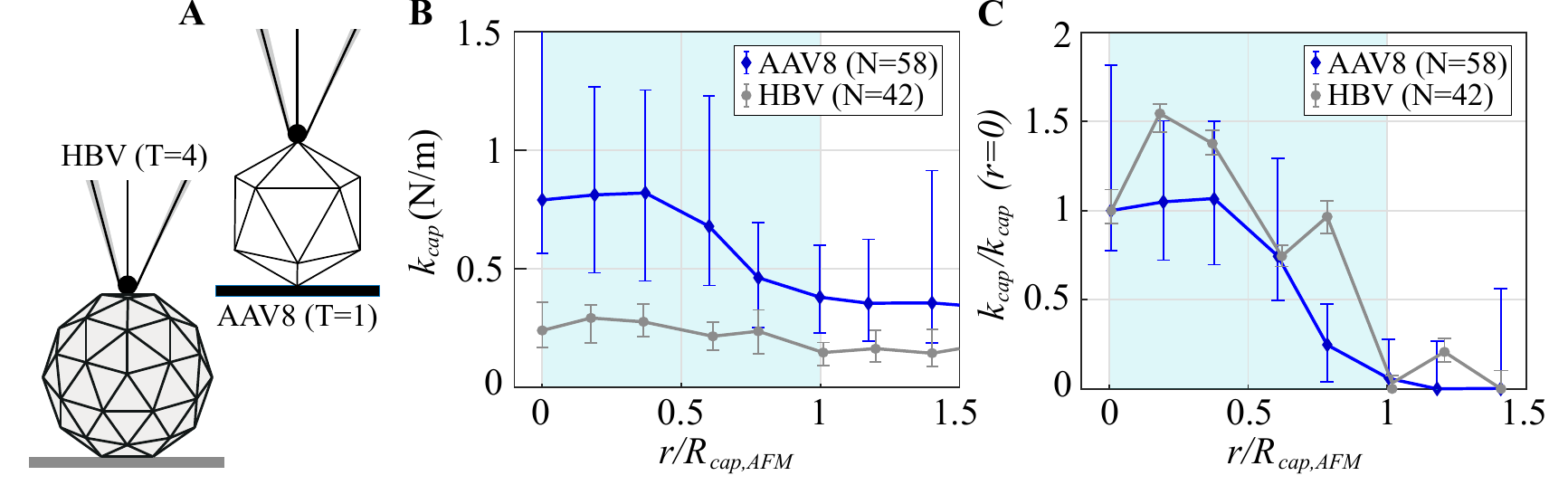}
\caption{\label{AAVHBV} Comparison between AAV8 and HBV average stiffness profiles. (\textbf{A}) Expected geometrical shape for AAV8 and HBV, based on their T-numbers. 
(\textbf{B}) Ensemble averaged stiffness profiles as function of lateral positions for AAV8 \emph{(blue curve)} and HBV \emph{(grey curve)}. For each position, the empty circle and extremities of the bar are respectively the median, the first and third quartile of the local stiffness distribution. (\textbf{C}) Rescaling of the two average profiles by their respective maximal amplitude, showing similar profiles}
\end{figure*}
First, we compared the average trajectory \textit{stiffness vs position} along AAV8 and HBV capsids measured for several viral particles (Fig.\ref{AAVHBV}B). For both capsids, 
the general trend is a decrease of stiffness from the center to the edge of the particle, in agreement with the analytical and numerical model. It is possible to rescale the stiffness with the top value and the position with the appropriate size $R+R_t$, so that both profiles collapse onto a single curve, up to the experimental error (Fig. \ref{AAVHBV}C). This is consistent with the geometrical modeling proposed so far. The median stiffness value on the top, $0.8N/m$ for AAV8 and $0.3N/m$ for HBV, is also consistent with other values from the literature \cite{zeng2017_2,roos2010,uetrecht2008}. 
Interestingly, we observed that the trajectories are relatively spread around the average trajectory, the effect being larger for AAV8 than HBV capsids, as  can be noticed from the values of the first and third quartile of the local stiffness distribution. 
\begin{figure*}[ht]
\centering
\includegraphics[width=15cm]{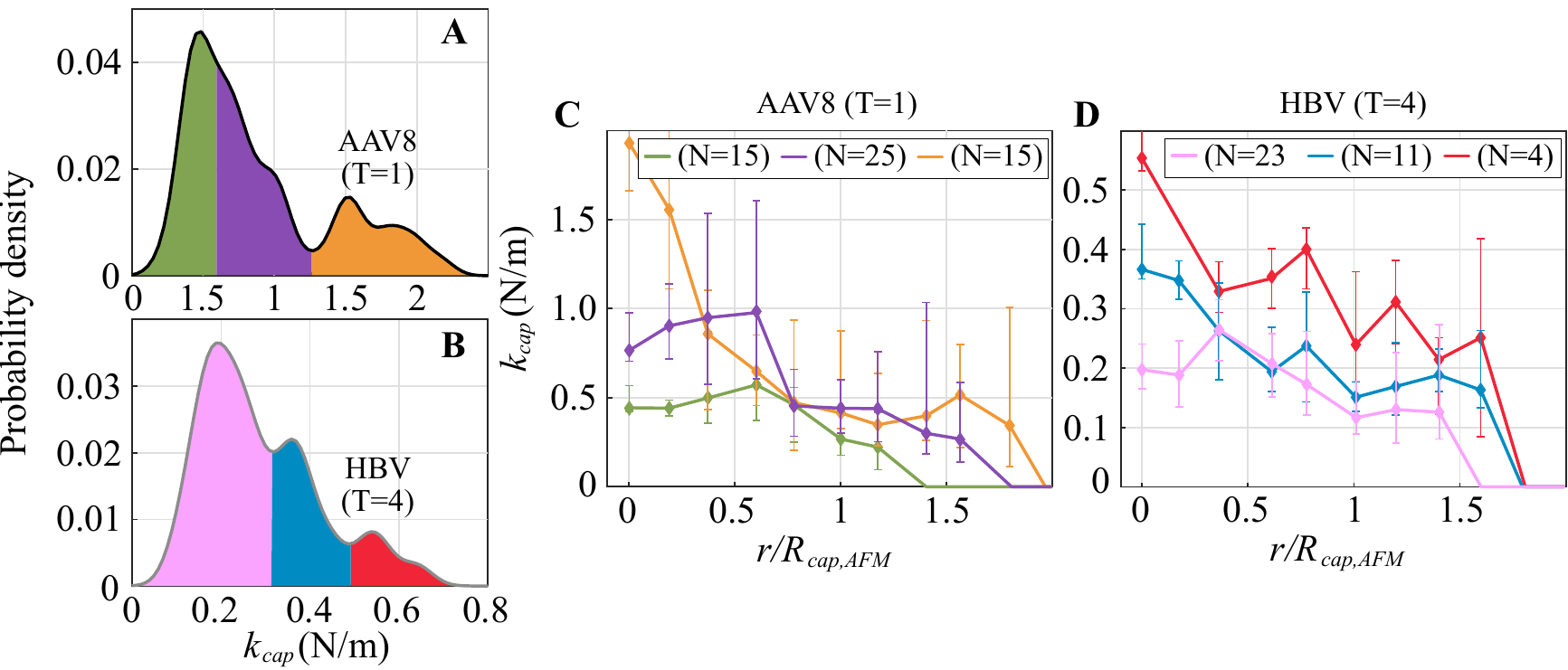}
\caption{\label{ProfileAAVHBV}Heterogeneity of stiffness profiles in the experiments. Stiffness distribution on the top of AAV8 (\textbf{A}) and HBV (\textbf{B}) viral capsids. Different subpopulations (identified by different colors) have been identified using slope ruptures in the cumulative top stiffness distributions for each curve. (\textbf{C}) Subpopulations of AAV8 stiffness profile with similar colors as in (\textbf{A}).  (\textbf{D}) Subpopulations of HBV stiffness profile with similar colors as in (\textbf{B}).}
\end{figure*}
In order to relate these results with the ones obtained in the simulation (fig.\ref{FvKInfluence}B), we extracted first the stiffness distribution on the top of the viruses for AAV8 and HBV (fig.\ref{ProfileAAVHBV}A and \ref{ProfileAAVHBV}B). These distributions show multimodal features. It is possible to use these modes of top stiffness distribution to select subpopulations of profile (fig.\ref{ProfileAAVHBV}C and \ref{ProfileAAVHBV}D). In practice, this is performed by inspecting the cumulative top stiffness distribution, which shows several significant slope ruptures see supplementary figure 4. These subpopulation profiles are clearly distinguishable, and they strongly suggest the presence of stiffness inhomogeneity. Now, within the simulations, the exact orientation of the adsorbed particles is known, and therefore it is likely that the observed subpopulations of profiles in the experiment are representative of the very same feature: the difference in orientation of adsorbed viral particles generates different stiffness profiles. Using subpopulation profiles, we recover the observations made in figure \ref{AAVHBV}B: relative inhomogeneity is stronger for AAV8 than for HBV capsids. Moreover, since the simulation suggests that heterogeneous stiffness profiles correlate to a faceted shape, we confirm from our mechanical experiment that stiffness heterogeneity of AAV8 viral capsids is associated with a more faceted shape than HBV viral capsid. This conclusion is in agreement with the structural information on these viral capsids, as AAV8 is a T=1 shell (strongly faceted), and HBV is a T=4 shell (weakly faceted, more spherical) \cite{wynne1999,roos2010,yu2013}.

\section{Conclusion}
In the present work, we investigated the sources of stiffness heterogeneities that can arise in indentation experiments of nano-sized shells like viral capsids. We first identified and rationalized a purely geometrical effect that reduces the stiffness of the shell as the vertical indentation axis is shifted towards the border of the shell. This is due to the decrease of the net deformation for identical vertical indentation. When compared to simulations, the prediction of a geometrical model that we proposed shows additionally that the adhesion properties of the shell to the substrate are rather important to account quantitatively for the observations. To our knowledge, this geometric effect has been already observed on influenza virus\cite{li2011}, but
this is the first time that it has been quantitatively investigated. It has some important consequences in the interpretation of nanoindentation experiments, as the stiffness value might be under-estimated in the case of measurements with poor positioning precision.

The recognition of this geometrical effect allowed us to define a new procedure to highlight intrinsic stiffness heterogeneity coming from the structural and elastic properties of the shell, and not from the relative geometry between the tip and the shell. Within this procedure, heterogeneity is evidenced by comparing stiffness trajectories obtained by performing several indentation experiments while scanning the investigated shell. Using simulations, we showed that if the shell is rather spherical, all trajectories collapse onto the expected geometrical profile. On the contrary, for a faceted shell resembling an icosahedron, stiffness trajectories are rather dispersed, reflecting intrinsic stiffness heterogeneity. We measured this effect experimentally  for two examples of small viruses, namely AAV8 and HBV (T=4). The different degree of intrinsic stiffness heterogeneity found for these viruses are consistent with their structure: AAV8 is a T=1 shell, with strong faceting, while HBV capsid is a T=4 with weaker faceting (more spherical shape); accordingly, the dispersion of stiffness profiles is larger for AAV8 than for HBV.

We acknowledge useful discussions at various stages of this work with Guillaume Tresset, Arezki Boudaoud and Catherine Quilliet. Part of this work has been funded through a CNRS "PEPS" grant entitled \textit{Modeling viral infections} in 2018. Y.C.-S. acknowledges the financial support of the National Agency for Research and Development (ANID) /  Scholarship Program / DOCTORADO BECAS CHILE 72170177. LL. acknowledges the financial support from the CNRS (CNRS-Momentum 2018).

\bibliographystyle{unsrt}
\bibliography{biblioINDENTATION}
\end{document}